\documentclass[a4paper,12pt]{article}

\usepackage{graphicx}
\usepackage[latin1]{inputenc}
\usepackage{amssymb}
\usepackage{amsmath}
\usepackage{epsfig}
\usepackage[margin=2.5cm,nohead]{geometry}

\newcommand{\ie}{{\it i.e.}}
\newcommand{\eg}{{\it e.g.}}

\newcommand{\pythia}{{\sc Pythia}}
\newcommand{\herwig}{{\sc Herwig}}

\newcommand{\bbar}{\bar b}

\newcommand{\pt}[1]{p_{T, #1}}
\newcommand{\mt}[1]{m_{T, #1}}
\newcommand{\GeV}{\mathrm{\;GeV}}

\title{An improved description of 
charged Higgs boson production\footnote{Talk given at the 42nd International School of Subnuclear Physics at Erice, Sicily, 1 September 2004}}
\author{Johan Alwall\\ 
\normalsize High Energy Physics, Uppsala Univ., Box 535, S-751 21 Uppsala, Sweden\\
\normalsize E-mail: {\tt Johan.Alwall@tsl.uu.se}}
\date{\today}

\begin{document}

\maketitle
\noindent{\bf Abstract:}\\
Many extensions of the Standard Model predict the existence of charged
Higgs bosons. In order to be able to find those particles, an accurate
description of their production is needed. In Monte Carlo simulations
of charged Higgs boson production at hadron colliders, the two
tree-level processes $gb\to H^\pm t$ and $gg\to H^\pm tb$ are
used. Since those processes overlap in the collinear region of the
phase-space of the outgoing $b$-quark, care must be taken not to
introduce double-counting if both processes are to be used
together. In this talk I present a method for matching these
processes, developed by Johan Rathsman and myself. The method also
allows for investigations of the factorization scale dependence of the
processes and a better understanding of which factorization scale to
choose to get a reliable description of charged Higgs production.


\section{Introduction}

The existence of a charged Higgs boson is a common feature of many
extensions of the Standard Model of particle physics, most notably
supersymmetric extensions such as the MSSM. In the Standard Model, the
fermions get their mass from their interaction with the Higgs field,
which gets a non-zero vacuum expectation value from spontaneous
symmetry breaking of the isospin $SU(2)_L$ symmetry. The vector bosons
$W^\pm$ and $Z^0$ get mass by absorbing three of the originally four
scalar degrees of freedom from the Higgs doublet. In supersymmetric
extensions of the Standard Model, one Higgs doublet is not enough - at
least two are necessary for mainly two reasons (see \eg\
\cite{Dawson:1996cq}):
\begin{enumerate}
\item The fermionic superpartner of the Higgs boson, so called higgsino, 
destroys the cancellation of the ABJ anomaly (see \eg\ \cite{Cheng}) in
the Standard Model. In order for the anomaly to cancel, two Higgs
doublets (and thus two higgsinos) with opposite hypercharge are
needed.
\item With supersymmetry, the same Higgs doublet cannot interact with
(and hence give mass to) both the up-type ($u$,$c$,$t$) and the
down-type fermions ($d$,$s$,$b$ and the charged leptons).
\end{enumerate}
With two Higgs doublets (8 real fields) but still only three massive
vector fields, we get five surviving Higgs fields:
\begin{center}
$h$, $H$, $H^+$, $H^-$, $A$ (pseudoscalar)
\end{center}
Such a theory is called a (type II) two Higgs doublet model (2HDM).

In the MSSM there are only two parameters determining the masses and
interactions of these fields: the first is
$\tan\beta=\frac{v_1}{v_2}$, the ratio of the vacuum expectation
values for the two Higgs doublets, the second is one of the masses, \eg\ the
pseudoscalar mass $M_A$. In a general 2HDM, however, there are seven
(or more) parameters.

\begin{figure}
\begin{center}
\epsfig{file=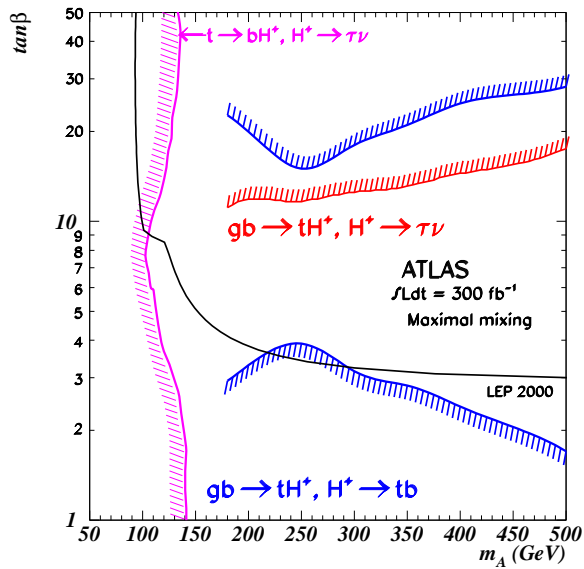,width=8cm}
\caption{\label{fig:region} The ATLAS 5-$\sigma$ discovery contour for charged Higgs. The gap in the region around the top mass could be bridged using a properly matched sum of the $gb\to H^\pm t$ process and the $gg\to H^\pm t b$ process. Figure taken from \cite{Assamagan:2002hz}.}
\end{center}
\end{figure}

Needless to say, the discovery of a charged scalar particle would be a
clear signal of physics beyond the Standard Model. In order to
search for such a particle, we need an accurate description of the
production mechanisms and phase-space distributions. Using Monte Carlo
programs, the production of charged Higgs can then be simulated, and
one can optimize search strategies (\ie\ minimize the Standard Model
background) using different cuts on the data from the collider. Even
before any actual experiment is done, one can in this way put limits
on the parameter-space regions where different experiments will be
able to find a signal (see fig.~\ref{fig:region}).

In Monte Carlo generators such as \pythia\cite{Pythia} and
\herwig\cite{Herwig}, the production channels used to simulate single 
charged Higgs boson production (as opposed to pair-production, which
is not discussed here) are $g\bar b\to H^+ \bar t$ and $gg\to H^+ \bar
t b $ and their charge conjugates. (There is also a process $q\bar
q\to H^\pm t b$, which gives a large contribution in a $p\bar
p$-collider such as Tevatron, but a very small contribution at the
high energies of the LHC.) Here (as usual) $g$ stands for gluon and
$q$ ($\bar q$) for an arbitrary quark (antiquark). The $gg\to H^\pm
tb$ process gives a better description of the part of phase-space
where the outgoing $b$-quark has a large transverse momentum ($p_T$),
while the $gb\to H^\pm t$ process resums potentially large logarithms
$(\alpha_s\log(\mu_F/m_b))^n$ and hence give a better description in
the rest of the phase space, as will be discussed later. In the region
where the outgoing $b$-quark has small transverse momentum, the two
processes overlap. Therefore, if both processes would be used and
summed naively we would get double-counting in this region of
phase-space. Together with Johan Rathsman, I have developed a method
to remove this double-counting by generating events from a
distribution corresponding to the double-counted part of phase space,
and subtracting these events from the sum. Our work is presented in
\cite{Alwall:2004xw}, where also more references are found.

\section{The twin-processes and their double-counting}

\begin{figure}
\begin{center}
\epsfig{file=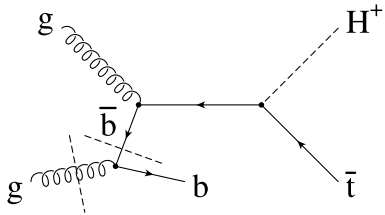,width=5cm}
\caption{\label{fig:overlap} Illustration of the relation between the  
$gb\to H^\pm t$ and $gg\to H^\pm tb$ processes. If the factorization
between the parton densities and the hard scattering is done at the
gluon line we get the $gg\to H^+ \bar t b$ process, while if instead
this factorization is done at the $\bar b$ line, we get the $g\bar
b\to H^+\bar t$ process. They can therefore be viewed as the same
process in two different approximations.}
\end{center}
\end{figure}

As discussed in the introduction, the two tree-level processes (\ie\
no-loop processes) used in Monte Carlo simulation of single production
of charged Higgs at hadron colliders are
\begin{eqnarray}
g\bbar(b) &\to& \bar t H^+ \; (t H^-) \label{eq:LO} \\
gg   &\to& \bar t b H^+ \;(t \bar b H^-) \label{eq:2to3}
\end{eqnarray}
The first one (\ref{eq:LO}), which I will denote the leading order
(LO) or $2\to2$ process, includes the $b$-quark density,
$b(\mu_F^2)\sim \sum (\alpha_s\log(\mu_F/m_b))^n$, which comes from
the logarithmic DGLAP resummation of gluon splitting to $b\bbar$
pairs. This means that the $b$-quark going into the process is
accompanied by a $\bar b$ (or vice versa) which is not explicitly
shown in the equation. Due to the approximation made in the DGLAP
expansion, this accompanying $b$-quark is nearly collinear with the
beam. The second production process, (\ref{eq:2to3}), which I will
denote the $2\to3$ process, gives the correct treatment of the
kinematics of the accompanying $b$-quark to order $\alpha_s^2$. The
relation between the two processes is illustrated in
fig.~\ref{fig:overlap}. Since the processes really have the same
initial and final states, they can be viewed as the same process in
two different approximations, hence the term ``twin-processes''.

As suggested by fig.~\ref{fig:overlap}, there is an overlap between
the two processes: When the transverse momentum of the outgoing
$b$-quark in the $2\to3$ process is small, there is no distinction
between the full $2\to3$ matrix element and a gluon splitting to
$b\bar b$ convoluted with the $gb\to H^\pm t$ matrix
element. Therefore there is a double-counting between the processes,
which can be expressed as \cite{Borzumati:1999th}
\begin{equation}
\label{eq:DC}
\sigma_\mathrm{DC}=\int dx_1dx_2\left[g(x_1,\mu_F)b'(x_2,\mu_F)
\frac{d\hat{\sigma}_{2\to 2}}{dx_1dx_2}(x_1,x_2) 
+ x_1 \leftrightarrow x_2\right]
\end{equation}
where
\begin{eqnarray} 
 \label{eq:bprime}
 b^\prime(x, \mu_F^2)
 &=&
 \frac{\alpha_s(\mu_R^2)}{2\pi}
 \int\frac{d Q^2}{Q^2+m_b^2}
 \int\frac{dz}{z} 
 P_{g\to q\bar q}(z) \; g\left(\frac{x}{z},Q^2\right) \\ 
 &\approx&
 \frac{\alpha_s(\mu_R^2)}{2\pi}\log\frac{\mu_F^2}{m_b^2}\int
 \frac{dz}{z} P_{g\to q\bar q}(z) \; g\left(\frac{x}{z},\mu_F^2\right)
\end{eqnarray}

This is just the leading logarithmic contribution to the $b$-quark
density included in the $2\to2$ process. Here $P_{g\to q\bar
q}(z)=\frac{1}{2}\left[z^2+(1-z)^2\right]$ is the splitting function
for $g$ going to $q\bar q$, $\mu_F$ is the factorization scale (\ie\
the scale where the parton densities are evaluated) and $\mu_R$ is the
renormalization scale used in evaluating $\alpha_s$, and $Q^2=-k^2$, the
4-momentum of the incoming $b$-quark squared. We need to take care
to include kinematic constraints due to the non-zero $b$-quark mass in
our calculation of the integration limits, since such constraints are
implicitly included in the $2\to3$ matrix element. This is done in
detail in our paper \cite{Alwall:2004xw}.

The matched integrated cross-section is then given by
\begin{equation}
\sigma = \sigma_{2\to 2} +\sigma_{2\to 3} \, -\sigma_\mathrm{DC}
\label{eq:xsec}
\end{equation}

\begin{figure}
\begin{center}
\epsfig{file=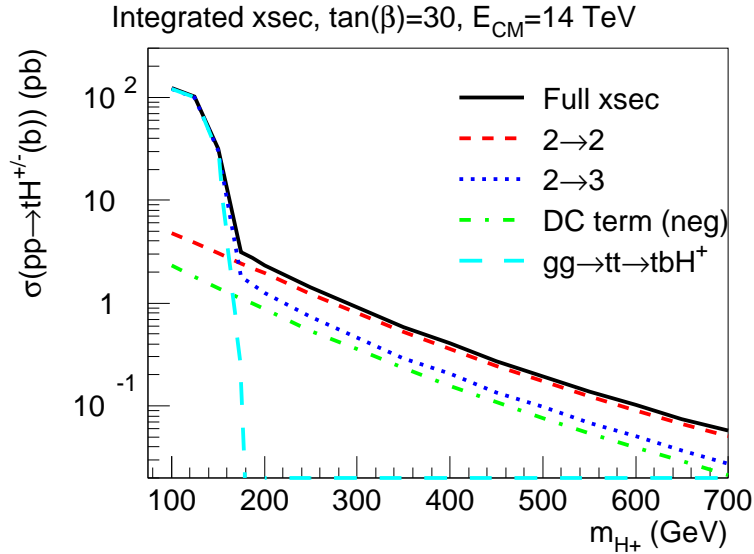,width=10cm}
\caption {\label{fig:massxsecs} Integrated cross-section components 
(leading order process, $2\to 3$ process and double-counting term) and
matched total as a function of the $H^\pm$ mass at LHC, with
$\tan\beta=30$ and $\mu_F=(m_t+m_{H^\pm})/4$. Note that the
double-counting term contribution (DC) is subtracted from the sum. At
$m_{H^\pm}<m_t$ the $2\to3$ process can be approximated by $gg\to
t\bar t\to tbH^\pm$.}
\end{center}
\end{figure}

The matched integrated cross-section and its components are shown as a
function of the charged Higgs boson mass in
fig.~\ref{fig:massxsecs}. For charged Higgs masses below the top mass
the cross-section can be well approximated by top pair production with
subsequent decay of one of the top quarks to $H^\pm b$ ,
\ie\ $gg\to t\bar t\to t b H^\pm$ (for a comparison between this
process and the $2\to3$ process, see \cite{Alwall:2003tc}). Our
matching procedure works for all charged Higgs masses, but is of
greatest interest for $m_{H^\pm}\gtrsim m_t$. In the following I will
use $m_{H^\pm}=250\GeV$ and $\tan\beta=30$ as a case study.

\section{Matching the differential cross-sections}

As we saw in the last section, the cancellation of the double-counting
between the $2\to2$ and $2\to3$ processes on the integrated
cross-section level was simple enough. But how do we do it for the
differential cross-sections? Whatever approach we take, we need to
make sure some basic requirements are fulfilled:

\begin{enumerate}
\item The integrated cross-section should equal the correct one given
by eq.~\eqref{eq:xsec}.
\item All differential cross-sections should be smooth after matching.
\item The matched $p_T$-distribution for the outgoing $b$-quark 
should be given by the $2\to2$ process for small transverse momenta,
and by the $2\to3$ process for large transverse momenta, with a smooth
interpolation between those regions.
\end{enumerate}

A Monte Carlo event generator simulates collision events by picking
kinematical variables from a probability distribution given by the
folding of parton density functions with the matrix element of the
hard scattering. In this way the events will be distributed in all
kinematic variables according to the differential cross-section
distribution of the process in question. This is true for the $2\to2$
process as well as for the $2\to3$ process. The number of events
generated for each process is proportional to the integrated
cross-section of the process. In a process where we have an incoming
$b$ ($\bar b$), such as the $2\to2$ process, we get an accompanying
outgoing $\bar b$ ($b$) from the gluon splitting as explained
above. The kinematic distribution of this accompanying $\bar b$ is
given by the DGLAP evolution and generated through so called parton
showers (PS) (see \eg\ \cite{Pythia}).

\begin{figure}
\begin{center}
\epsfig{file=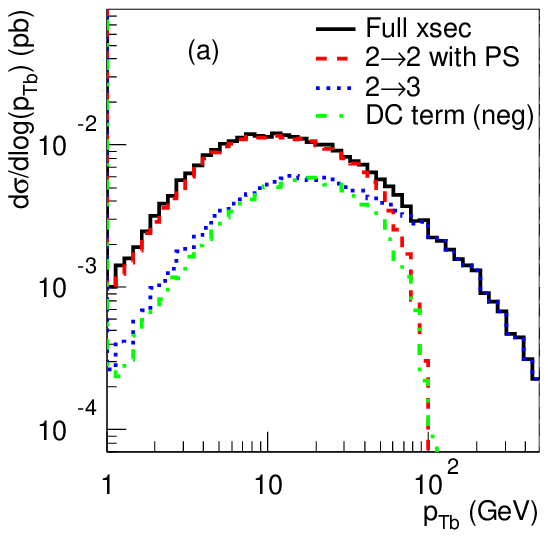,width=6.5cm}
\epsfig{file=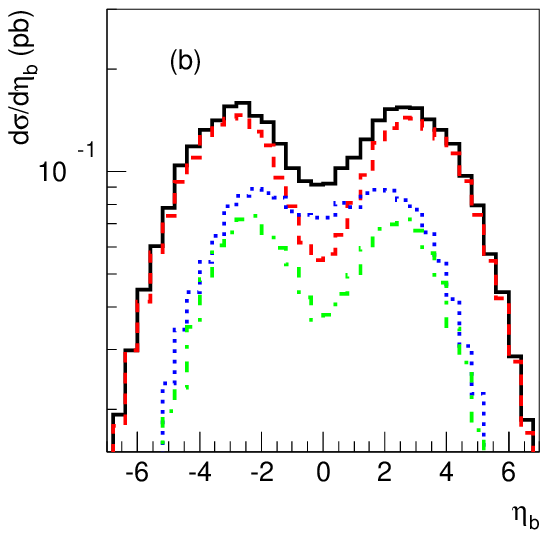,width=6.5cm}
\epsfig{file=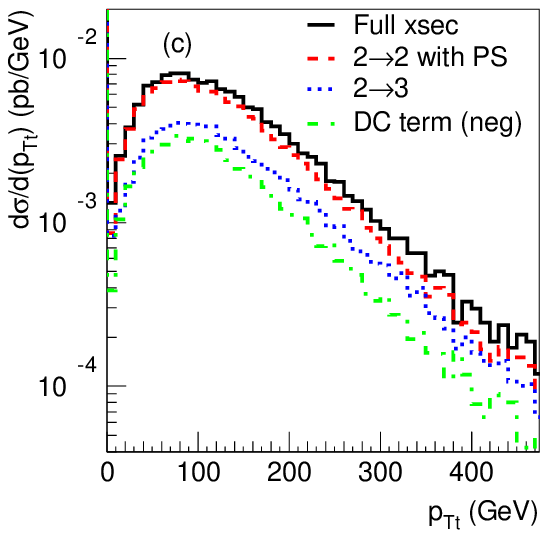,width=6.5cm}
\epsfig{file=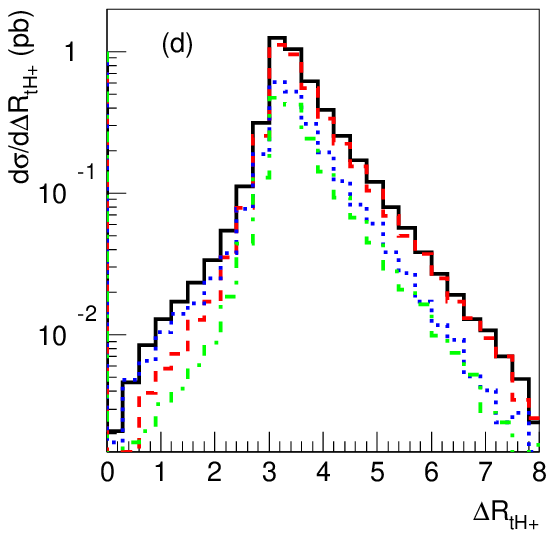,width=6.5cm}
\caption{\label{fig:diff-xsecs} Differential cross-sections in 
(a) $\pt{b}$, (b) $\eta_b$, (c) $\pt{t}$ and (d) $\Delta R_{H^\pm t}$ for 
the cross-section components and the resulting matched
cross-section with $\tan\beta=30$, $m_{H^\pm}=250\GeV$ and
$\mu_F=(m_t+m_{H^\pm})/4$. Note that the double-counting term
contribution (DC) is subtracted from the sum.}
\end{center}
\end{figure}

Now, our solution to the matching problem is simple: We view the
double-counting term (given by eq.~\eqref{eq:DC}) as a probability
distribution in kinematic variables and pick events from this
distribution. Then we subtract this contribution (\ie\ add with
negative weight) from the sum of the two processes \eqref{eq:LO} and
\eqref{eq:2to3} in the final data analysis, \ie\ the 
histograms.\footnote{The code for the double-counting term,
implemented as an external process to \pythia, will be available
for download at {\tt http://www3.tsl.uu.se/thep/MC/pytbh}, including a
manual in preparation.}

One might worry that this procedure could give a negative number of
events in some phase-space region. However, the leading-logarithmic
part of the $b$ density used in the double-counting term is always
smaller then the full $b$ density used in the $2\to2$ process,
ensuring that if only a sufficiently large number of events is
generated, the sum of the events from the $2\to2$ and the $2\to3$
process will always be larger than the number of events from the
double-counting term.

Some resulting differential cross-sections from our matching are shown
in fig.~\ref{fig:diff-xsecs}. Looking at fig.~\ref{fig:diff-xsecs}a we
see that the matched differential cross-section in $p_T$ for the
outgoing $b$-quark looks exactly as expected (and wanted): for small
$p_T$ it follows the $2\to 2$ process distribution, while for large
$p_T$ it follows the $2\to3$ process. However, there is a rather large
intermediate region, from about $30\GeV$ to about $100\GeV$, where the
whole matching procedure is really necessary to get the correct
differential cross-section. For comparison one can note that at LHC,
the region where $b$-quarks can be tagged is $\pt{b}\gtrsim
20\GeV$. In fig.~\ref{fig:diff-xsecs}b we see large
differences in the rapidity distribution for the outgoing $b$-quark
between the matched cross section and each of the two contributing
processes, up to a factor $\simeq 2$ in the experimentally interesting
region $|\eta_b|<2.5$, indicating that matching is really
necessary to get a correct description of the $b$-quark kinematics..

Even if the outgoing $b$ is not observed, we still see a difference
between the matched cross-section and the $2\to2$ process, which is
usually used in this case. In the transverse momentum distribution of
the top quark (fig.~\ref{fig:diff-xsecs}c) we get a small enhancement
for large $\pt{t}$, amounting to $\sim 20\%$ compared to the
$2\to2$ process. There is a similar effect in the distribution of the
charged Higgs boson (not shown). In fig.~\ref{fig:diff-xsecs}d, showing the
distance measure $\Delta R_{H^\pm t} =\sqrt{\Delta\varphi_{H^\pm
t}^2+\Delta\eta_{H^\pm t}^2}$, we see an effect especially for
small separations.

All in all, we see that unless we make very limiting cuts on the phase
space, we need to use the matched differential cross-sections to get
reliable predictions.

\section{Factorization scale dependence}

The factorization scale is the scale where the factorization is done
between the parton description of the proton and the hard scattering
of the partons, \ie\ where the parton distributions are
evaluated. This scale also determines the maximum transverse momentum
of the parton shower, \eg\ of the outgoing $b$-quark of the $2\to2$
process. Of course, the factorization scale, just like the
renormalization scale, is a fictitious entity in the sense that if all
orders of perturbation theory was included in the cross-section
calculation, the scale would never show up in the result. However, in
real life we can only use a few orders in perturbation theory, and
then it is necessary to introduce such scales.  Often the
factorization scale is chosen to be some hard scale of the process,
\eg\ the invariant mass $\hat s$ of the partonic system or the
transverse mass of one of the outgoing particles. In the process we
are discussing, $gb\to H^\pm t$, often $m_{H^\pm}+m_t
\approx \hat s$ is used as the factorization scale. What scale we should 
use in the lower-order calculations can not really be known until all
orders of perturbation theory have been calculated, but the
factorization scale dependence should be weaker the more orders we
include. Therefore we get an indication of which scale to use in the
first-order expression already by doing the next-to-leading order
calculation.

\begin{figure}
\begin{center}
\epsfig{file=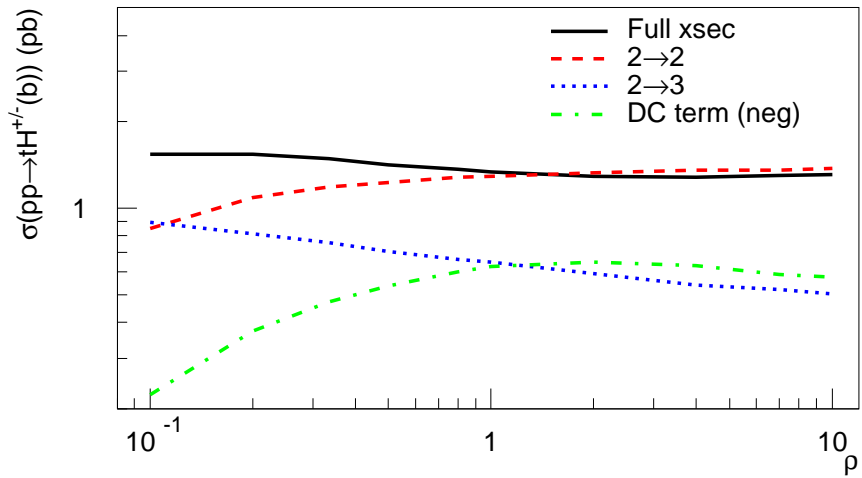,width=12cm}
\caption{\label{fig:rhoxsecs} Integrated cross-sections at $m_{H^\pm}=250\GeV$
and $\tan\beta=30$ as a function of the factorization scale
parametrized by $\rho=2\mu_F/(m_t+m_{H^\pm})$. Note that for
$\rho\gtrsim 1$ the double-counting term exceeds the $2\to3$
term. Also note that the factorization scale dependence is smaller for
the matched total cross-section than for any one of the components.}
\end{center}
\end{figure}

\begin{figure}
\begin{center}
\epsfig{file=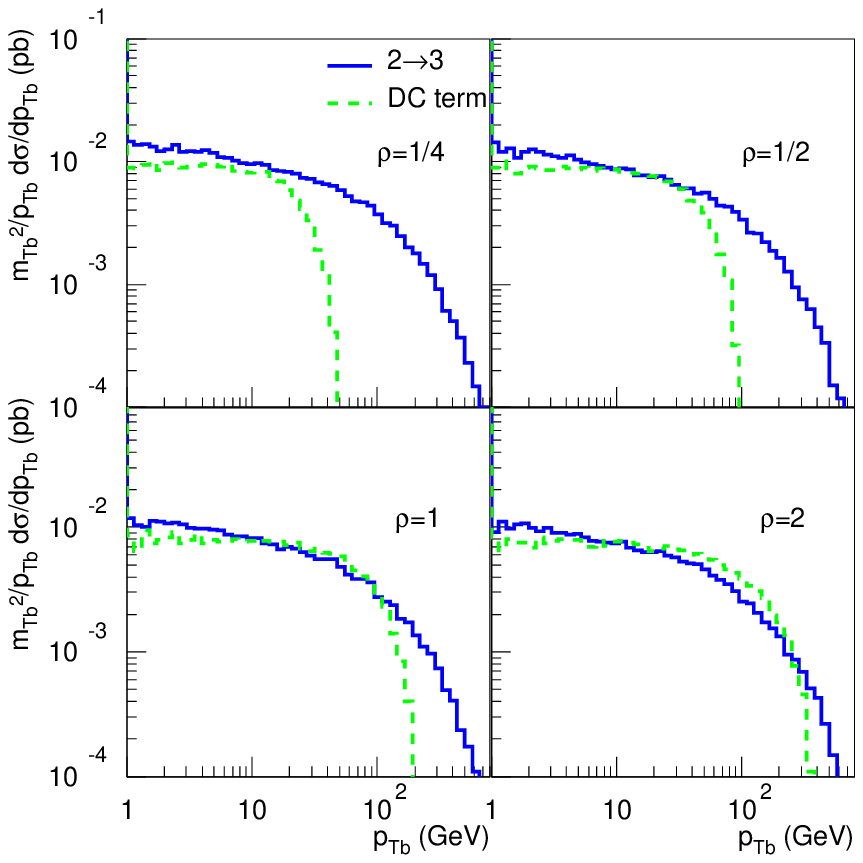,width=12cm}
\caption{\label{fig:plateau4} The differential cross-section $d\sigma/d\pt{b}$
multiplied by $\mt{b}^2/\pt{b}$ for the $2\to 3$ matrix element and
the double-counting term for different factorization scales
parameterized by $\rho=2\mu_F/(m_t+m_{H^\pm})$. Note that the
double-counting term overshoots the $2\to3$ term already for
$\rho=1$.}
\end{center}
\end{figure}

The double-counting term of eq.~\eqref{eq:DC} is designed to describe
the part of the $2\to3$ process which is already included in the $2\to
2$ process through the resummation of logarithmic contributions to the
$b$-quark density. This means that the double-counting term should not
exceed the $2\to3$ process contribution in any part of the
phase-space. However, the upper integration limit of the transverse
momentum of the outgoing $b$-quark in the double-counting term is
determined by the factorization scale. Up to this limit the transverse
momentum-distribution is almost flat, which means that the integrated
value of the double-counting term is nearly proportional to the
factorization scale (although for large scales the kinematic
constraints modify this behaviour). This is illustrated in
figs.~\ref{fig:rhoxsecs} and \ref{fig:plateau4}.

Fig.~\ref{fig:rhoxsecs} shows the integrated cross-section components,
as well as the matched total, as a function of the factorization scale
$\mu_F$ parametrized by $\rho=\mu_F/\overline m$, where $\overline
m=(m_{H^\pm}+m_t)/2$ is the average of the charged Higgs and top
masses. Here two things can be noted. The first is that the
double-counting term exceeds the $2\to3$ process term already near
$\rho=1$, \ie\ $\mu_F\gtrsim (m_{H^\pm}+m_t)/2$. This indicates that
the factorization scale should definitely not be chosen above this
value. The other thing to be noted is that the matched integrated
cross-section shows a significantly smaller dependence on the
factorization scale than any one of the component cross-sections,
indicating that the matching also stabilizes the cross-section. This
is not surprising, since the $2\to3$ process is a part of the
next-to-leading order calculation of the $2\to2$ process. Hence
including it in a correct way should reduce the factorization scale
dependence, as is the case when calculating the full
next-to-leading order expression.

In fig.~\ref{fig:plateau4} we see the distribution in $p_T$ of the
outgoing $b$-quark for the $2\to3$ process (without parton showers)
and the double-counting term at different factorization scales. Here
we see that already for $\rho=1$, the double-counting term exceeds the
$2\to3$ term around $\pt{b}\approx 50\GeV$. $\rho=0.5$, \ie\
$\mu_F=(m_{H^\pm}+m_t)/4$, seems to be a limiting case where the
double-counting distribution just touches the
$2\to3$-distribution. Therefore we have chosen to use this
unconventionally small value of the factorization scale in our plots
in the previous sections. Similar results for the size of the
factorization scale has been achieved in next-to-leading order
calculations of the $2\to2$ process, see
\cite{Plehn:2002vy,Berger:2003sm}.

\section{Conclusions}

The discovery of a charged scalar particle would be a clear signal of
physics beyond the Standard Model of particle physics. The detailed
study of the properties of such a particle would give
valuable insight into the nature of this new physics. But to be able
to find the charged Higgs boson, we need to devise search strategies
and to reduce the Standard Model background. For this, an appropriate
description of charged Higgs boson production, using Monte Carlo
events generators, is necessary.

For single charged Higgs boson production, mainly two processes are
used in Monte Carlos, the bottom-gluon fusion $gb\to H^\pm t$
production channel and the gluon-gluon fusion $gg\to H^\pm t b$
channel. These have different virtues: $gb\to H^\pm t$ resums large
logarithms describing the $b$-quark density, why it gives the major
contribution to the total cross-section, and also gives the best
description of the differential cross-section for small values of the
transverse momentum of the associated $b$-quark. On the other hand,
the $gg\to H^\pm t b$ process gives a correct description to order
$\alpha_s^2$ of the outgoing $b$-quark for large values of the
transverse momentum. However, the two processes overlap in the small
transverse momentum region, so in order to use them both we must
compensate for this double-counting.

In this talk I have presented our algorithm for the matching of the
two processes. This matching is done by summing the events from the
two processes (as is usually done in Monte Carlo generators), and then
subtract events generated from a double-counting distribution term
from the sum. In this way we are able to combine the virtues of the
two processes to get a good description of the full differential
cross-section.

This method also allows us to get a better understanding of what
choice of factorization scale is appropriate, by comparing the
transverse momentum distributions for the double-counting term with
the distribution for the $2\to3$ process matrix element. Since the
double-counting term should remove the part of the $2\to3$
distribution already contained in the $2\to2$ distribution, the
double-counting term should not overshoot the $2\to3$ term. The result
is that the appropriate factorization scale is significantly smaller
than the conventionally used value.
\\

\noindent{\bf Acknowledgements:}\\ I would like to thank the organizers of
ISSP42 for giving me the opportunity to present this talk. I would
also like to thank all participants of the school for giving me a
great time.

\end{document}